\documentclass[conference]{IEEEtran}
\IEEEoverridecommandlockouts
\usepackage{cite}
\usepackage{amsmath,amssymb,amsfonts}
\usepackage{algorithmic}
\usepackage{graphicx}
\usepackage{textcomp}
\usepackage{xcolor}
\usepackage{booktabs}
\usepackage{pdflscape}
\usepackage{hyperref}
\usepackage{graphicx}
\usepackage{tabularx}
\usepackage{hyperref}
\usepackage{graphicx}
\usepackage{placeins} % For \FloatBarrier command
\usepackage{lipsum} % For dummy text

\usepackage{float}

\usepackage{cite}
\usepackage{multirow}

\usepackage{tabularx}
\def\BibTeX{{\rm B\kern-.05em{\sc i\kern-.025em b}\kern-.08em
    T\kern-.1667em\lower.7ex\hbox{E}\kern-.125emX}}
\begin{document}

% \title{
% IMPACT FALL DETECTION  USING MACHINE LEARNING}
\title{ Machine Learning and Feature Ranking for Impact Fall Detection Event Using Multisensor Data}

\author{\IEEEauthorblockN{\textsuperscript{} Tresor Y. Koffi}
\IEEEauthorblockA{\textit{CESI LINEACT Laboratory, } \\
\textit{UR 7527}\\
Dijon, 21800, France
 \\
ytkoffi@cesi.fr}
\and
\IEEEauthorblockN{\textsuperscript{} Youssef Mourchid}
\IEEEauthorblockA{\textit{CESI LINEACT Laboratory, } \\
\textit{UR 7527}\\
Dijon, 21800, France
 \\
ymourchid@cesi.fr\\ 
\and
\IEEEauthorblockN{\textsuperscript{} Mohammed Hindawi }
\IEEEauthorblockA{\textit{CESI LINEACT Laboratory, } \\
\textit{UR 7527}\\
Lyon, 69100, France \\
mhindawi@cesi.fr}
}
\and

\IEEEauthorblockN{\textsuperscript{} Yohan Dupuis }
\IEEEauthorblockA{\textit{CESI LINEACT Laboratory, } \\
\textit{UR 7527}\\
Paris La Défense, 92074, France
 \\
ydupuis@cesi.fr}
}

\maketitle
\begin{sloppypar}
\begin{abstract}
Falls among individuals, especially the elderly population, can lead to serious injuries and complications. Detecting impact moments within a fall event is crucial for providing timely assistance and minimizing the negative consequences. In this work, we aim to address this challenge by applying thorough preprocessing techniques to the multisensor dataset, the goal is to eliminate noise and improve data quality. Furthermore, we employ a feature selection process to identify the most relevant features derived from the multisensor UP-FALL dataset, which in turn will enhance the performance and efficiency of machine learning models.
We then evaluate the efficiency  of various machine learning models in detecting the impact moment using the resulting data information from multiple sensors. Through extensive experimentation, we assess the accuracy of our approach using various evaluation metrics. Our results achieve high accuracy rates in impact detection, showcasing the power of leveraging multisensor data for fall detection tasks. This highlights the potential of our approach to enhance fall detection systems and improve the overall safety and well-being of individuals at risk of falls.

\end{abstract}


\section{Introduction}\label{sec:introduction}

According to the World Health Organization report, falls affect 32 percent of older adults annually, making it a leading cause of mortality among this demographic \cite{kundakcci2020determination}. The elderly population is particularly vulnerable to the detrimental effects of falls, which can result in fractures and cognitive impairments\cite{karlsson2013prevention}. Recent research efforts have focused on fall detection to predict and prevent such incidents \cite{fahimi2021identifying}. The primary objective of these studies is to minimize fall-related injuries and provide timely assistance to individuals at risk. An important advancement in this field is the implementation of pre-impact fall detection systems\cite{chi2023prefallkd}. These systems identify falls at an early stage before the impact occurs.

\begin{figure}[t]
    \centering
     \includegraphics[width=0.45\textwidth, height=0.4\textheight]{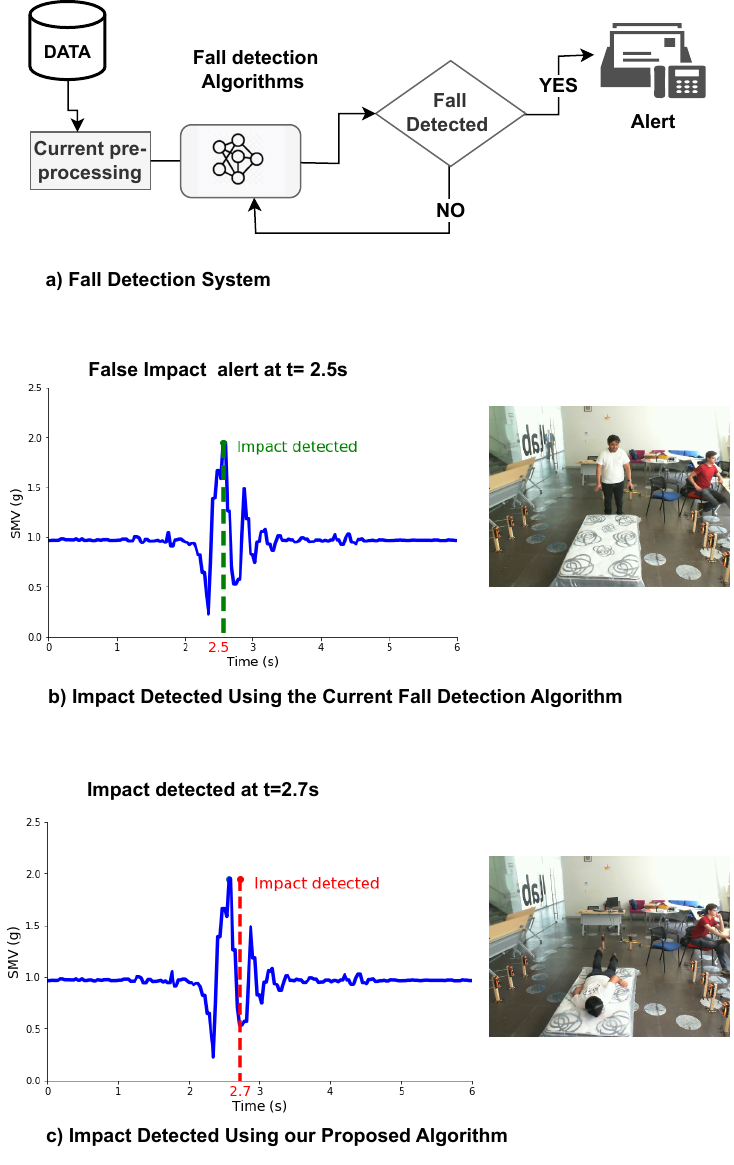}
    \caption{Comparison of impact detection using the state of the art method  with our proposed method}
    \label{impact}
\end{figure}

Advancements in computer vision, driven by graphs, statistical techniques, and deep learning, have greatly enhanced visual data processing, which is particularly beneficial for improving fall detection accuracy \cite{mourchid2016image,benallal2022new,mourchid2021automatic,mourchid2023mr}.
To detect falls across different stages, including pre-impact, and post-impact, accelerometer-based fall detection systems have gained prominence.
However, while identifying fall at the early stage and post-stage can help to rescue people, it is crucial to identify the impact point, in order to avoid false alert and accurately provide assistance to people who really fall. Furthermore, it has been observed that the current datasets available for fall detection lack proper preprocessing, leading to inaccurate detection of falls and impacts in real-life scenarios. Therefore, depending on raw data from these datasets for the development of fall detection systems can lead to an increased occurrence of false positives or false negatives, as illustrated in Fig. \ref{impact}(b). This figure demonstrates the occurrence of a false positive, where the system incorrectly identifies an impact where it has not actually occurred. This can cause false alarms and that reduces trust in the detection system.  False negative cases are more dangerous in this context, where the person has actually fallen but the system does not detect it. Moreover, the detection of impacts from these datasets may suffer from delays, which further undermines the reliability of fall detection systems as shown in Fig. \ref{impact}(a).

To address these limitations, we emphasize the importance of preprocessing techniques. Furthermore, a feature selection method is employed  to identify the relevant features from the dataset. This method ranks the features based on their importance and relevance to the task of fall and impact detection. Hence, the main contributions of this paper are as follows:
\begin{enumerate}
  \item A preprocessing technique for a dataset is implemented to ensure that fall detection and impact detection algorithms operate on clean and accurate data, enabling more robust and precise results.
  
  \item A feature selection process is employed which aims to identify the most relevant and informative features for impact detection, with the objective of reducing dataset dimensionality and focusing on the features that significantly contribute to accurate impact detection.
  
  \item Machine learning approaches are applied to precisely identify the impact points with the ground during a fall event. This ensures that only genuine falls are detected, and the person has actually made contact with the ground.

\end{enumerate}\
The rest of the paper is organized as follows. \hyperref[sec:related_work]{Section II} presents a review of the related works in fall detection and impact detection. In \hyperref[sec:methodology]{Section III}, we discuss in detail the methodology employed in this study. It includes the description of the fall dataset, the preprocessing techniques applied, the feature selection process, and the machine learning algorithms utilized for impact detection.
\hyperref[sec:results]{Section IV}
and \hyperref[sec:conclusion]{Section V}
 present respectively the results obtained from the experiments conducted in this study, the conclusion, and the perspectives.
 
 % \FloatBarrier

% \begin{figure}[t]
%     \centering
%     \includegraphics[width=0.6\textwidth, height=0.4\textheight]{figures/falldetection_figure.pdf}
%     \caption{Comparison of impact detection using the state-of-the-art method with our proposed method}
%     \label{impact}
% \end{figure}

\section{Related Works}\label{sec:related_work}

Falls among the elderly pose a significant health concern. In the population aged 65 and above, falls contribute to nearly half of all hospital admissions related to injuries\cite{miskelly2001assistive}. However, the real definition of a fall seems to be challenging, making it difficult to establish an accurate detection method. Thence, the threshold accelerometers rule is the most fall detection method used to detect falls. These rules are used to determine if a potential fall has occurred\cite{jantaraprim2009evaluation}. While accelerometer-based fall detection methods can be effective in detecting falls, there are limitations when it comes to accurately determining the impact time and providing precise assistance. The impact time refers to the exact moment when the individual makes contact with the ground, and it can be challenging to accurately determine this using accelerometer data alone. Factors such as the orientation of the sensor, and the type of fall can affect the reliability of impact time estimation. In this context, the fixed-size non-overlapping sliding window (FNSW) and fixed-size overlapping sliding window (FOSW) approaches have been used  to detect pre-impact and post-impact moments \cite{putra2017event}. These methods are typically applied during the pre-processing phase of fall detection system. However, it is  important to note that these methods do not provide real-time fall detection or accurate estimation of the impact moment. In addition to this approach, authors in  \cite{yu2020novel} proposed an architecture based on deep learning to detect the pre-impact phase in fall events. Deep learning  approaches learn complex patterns and temporal dependencies from the accelerometer data \cite{yu2020novel}. Their approach improved the accuracy  of pre-impact detection compared to traditional feature-based methods. However, the performance of deep learning models still depends heavily on the availability of labeled data and the quality of the training process.

\section{Methodology}\label{sec:methodology}

\begin{figure*}[t]
    \centering
    \includegraphics[width=16cm,height=4cm]{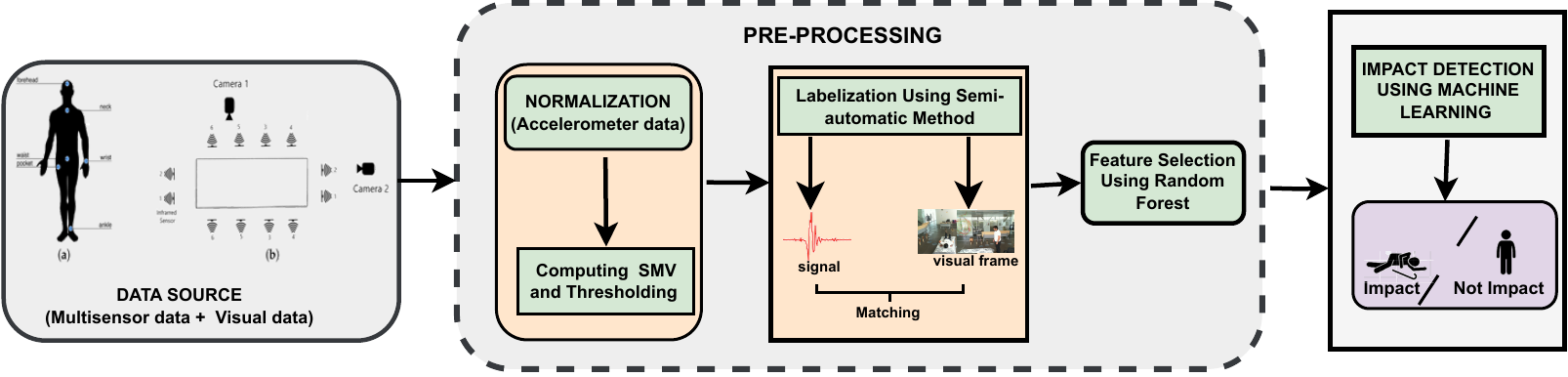}
    \caption{The Proposed Approach Pipeline}
    \label{pipline2}
\end{figure*}

In this section, we will present a detailed pipeline in Fig. \ref{pipline2} that highlights the steps involved in our methodology. This figure provides a visual representation of how the data preprocessing, feature selection, and machine learning models training are integrated to achieve accurate impact identification within fall events.

\subsection{DATASET}

Since obtaining real fall data from the elderly is still a real challenging task due to the privacy issue, for this study we  use the  UP-Fall dataset\cite{martinez2019up}. As a public dataset, it is specially designed for fall detection not impact detection purposes. It comprises a diverse range of activities and trials to facilitate comprehensive analysis. Thus, the dataset includes 11 distinct activities and six activities of daily living (ADLs) with each activity being performed three times. Furthermore, the dataset includes a range of many accelerometers, gyroscopes, EEG sensors and vision  devices used to capture visual data and provide a visual understanding of the environment.

\subsection{DATA PREPROCESSING}

UP-fall dataset \cite{martinez2019up} has been specifically curated and processed for the purpose of fall detection. Therefore the preprocessing methods applied to this dataset are tailored to enhance the reliability of fall detection algorithms\cite{al2021towards}. Since the objective of this work is to clearly identify impact point within fall event, we proceed to a further pre-processing of this dataset. The crucial part during our data preprocessing phase was to synchronize sensor data with the visual data. This step was necessary as we are dealing with multiple sensor data collected  from different accelerometers at different timestamps. Therefore, synchronizing the sensor data with the visual data involves matching the timestamps of the sensor readings with the corresponding visual frames. This alignment enabled a coherent understanding of the events captured by the sensors and the corresponding visual information.
However, as part of the data pre-processing, it is essential to apply a suitable normalization technique to ensure that all features are treated fairly and enable accurate assessment of their importance for impact identification in fall events.
Thus, to accurately detect the impact and tally with our research objective the following methods were used. 

\begin{itemize}
	
\item \textbf{Feature Normalization}\end{itemize}
The analysis of the UP-fall dataset revealed the presence of 37 features, some of which may contain special values or missing values (NAN). To ensure a fair and consistent comparison of these features, the Z-score normalization technique \cite{jeong2022fall} is used to scale all the features to the same range. This technique is employed to assess the efficiency of normalization and guarantee that all the features are on a standardized scale.
	
\begin{itemize}
\item \textbf{Signal Magnitude Vector-based Method}\end{itemize}

Signal Magnitude Vector (SMV) refers to the overall magnitude or amplitude of the accelerometer data. This method is used in the fall detection \cite{al2021towards}. It computes the Euclidian norm or the magnitude of the vector formed by the three axial accelerometers.

The SMV score can be calculated using the following formula:
\begin{equation}
\text{SMV} = \sqrt{A_x^2 + A_y^2 + A_z^2} \tag{1}
\end{equation}
where $A_x$, $A_y$, and $A_z$ represent the respective acceleration components along the $x$, $y$, and $z$ axes.

\begin{itemize}
\item \textbf{Threshold-based method}\end{itemize}
This method involves defining a threshold value that acts as a boundary of the cutoff point. Thus, based on the movement of the inertial force, the threshold-based approach is used to detect fall and not fall. Several research studies have used the threshold technique to detect fall onset\cite{putra2017event}. The fall is considered happening if the measured acceleration, denoted as \(a\), exceeds a predefined threshold.

The fall detection formula can be expressed as follows:

\begin{equation}
\lvert a \rvert > \beta \tag{2}
\end{equation}

where $\lvert a \rvert$ represents the absolute value of the measured acceleration, and $\beta$ represents the threshold value. Typically, the threshold for non-event is defined as \(1g\), where 
\(g\) represents the acceleration due to gravity. Therefore, any value  above this threshold is considered as the falling onset, as described in Fig. \ref{impact}(b).

Based on the aforementioned studies, accurate detection of fall events and the identification of impact points require a combination of preprocessing techniques to reduce noise in the dataset. In our approach, we will first calculate SMV of the accelerometer data. Simultaneously, we set a threshold value of \(2g\) to determine the presence of a fall event. The threshold value of \(2g\) is chosen based on several considerations. Firstly, we aim to filter out the noise and ensure that only significant accelerations associated with falls are considered. By setting a higher threshold, we can focus on capturing more pronounced and impactful accelerations.

\begin{figure}[t]
    \centering
    \includegraphics[width = 0.49\textwidth, height=12cm]{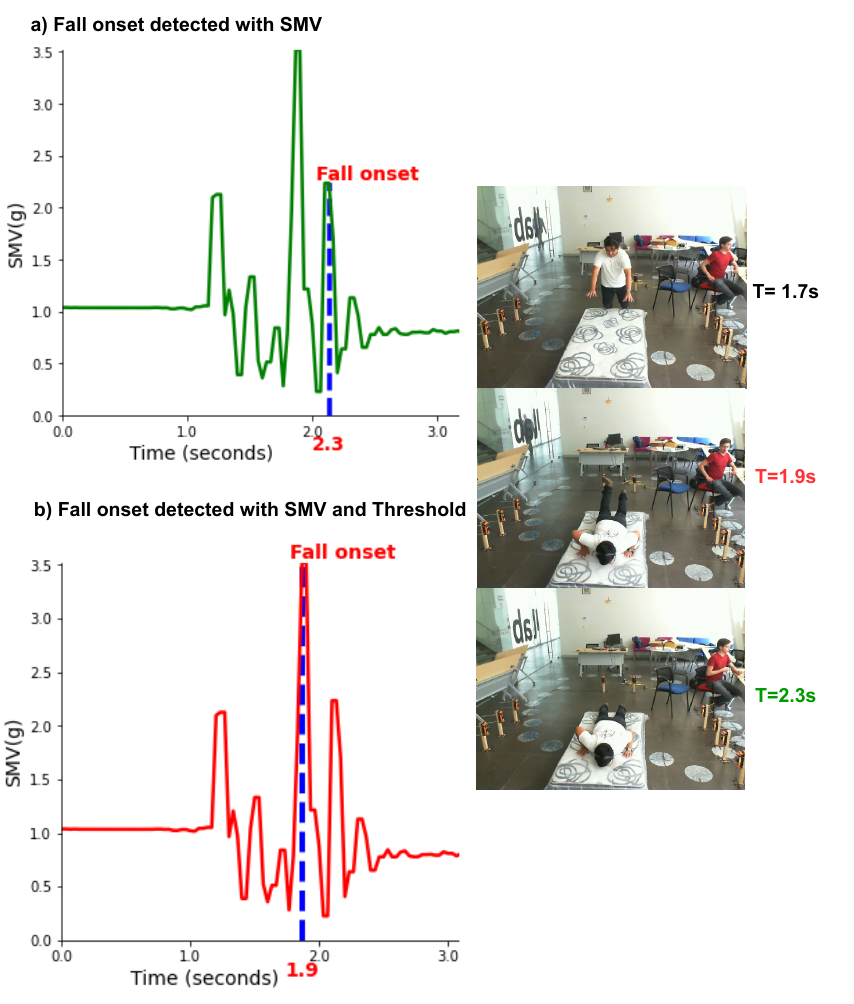}
    \caption{Fall onset detection : a)  with only SMV,  b) with SMV and Threshold}
    \label{fallonset}
\end{figure}

Furthermore, the choice of \(2g\) is influenced by previous research in the field. Many studies have used a threshold value of 1 to  \(1.6g\) to detect falls\cite{bourke2007evaluation}. However, when we visually checked the results by synchronizing them with the visual frames from the visual data, we observed that these threshold values did not accurately reflect the fall events or the impact points. Therefore, selecting a threshold above \(2g\) aims to ensure that the person has successfully made contact with the ground.
As shown in Fig. \ref{fallonset}(a) , the fall onset appears at t= 2.3 seconds meaning that applying only SMV method \cite{al2021towards} without a threshold will potentially lead to a delay or  false fall onset detection.
However, in Fig. \ref{fallonset}(b), when we apply the SMV calculation and set the threshold value simultaneously, the fall onset occurs at t=1.9 seconds, which tallies with the actual fall event. This indicates that our approach successfully reduces false positive detection and improves the accuracy of fall event detection which can lead to accurate impact detection. However, despite the use of this method, the ground impact detected sometimes not corresponding to the real impact point in the frame of the visual data. Therefore it is essential to proceed to further preprocessing.

\begin{itemize}

\item \textbf{Semi-automatic method for data labeling}\end{itemize}

This section provides a further contribution on the dataset to detect the impact within fall events. Thus, to reduce noise in our data and delay within fall  event detection, we compare the calculated SMV value with a predefined threshold for impact and not impact event as follows:
\begin{equation}
\begin{cases}
   \text{if } \text{SMV} > \beta, \quad \text{then impact detected} \\
   \text{if } \text{SMV} \leq \beta, \quad \text{then no impact detected}
\end{cases}
\tag{3}
\end{equation}
where $\beta$ represents our given threshold value.

In order to accurately label the data, the impact is considered detected if the SVM value exceeds the given threshold of \(2g\). Thus, to validate each detected event, we perform a visual check by examining the corresponding visual data. This process allows us to split the data and label the impact data as 1 and not impact data as 0. Furthermore, this validation process ensures that each event detected aligns with the corresponding frame detected in the visual data.

\begin{figure}[t]
    \centering
    \includegraphics[width = 0.5\textwidth]{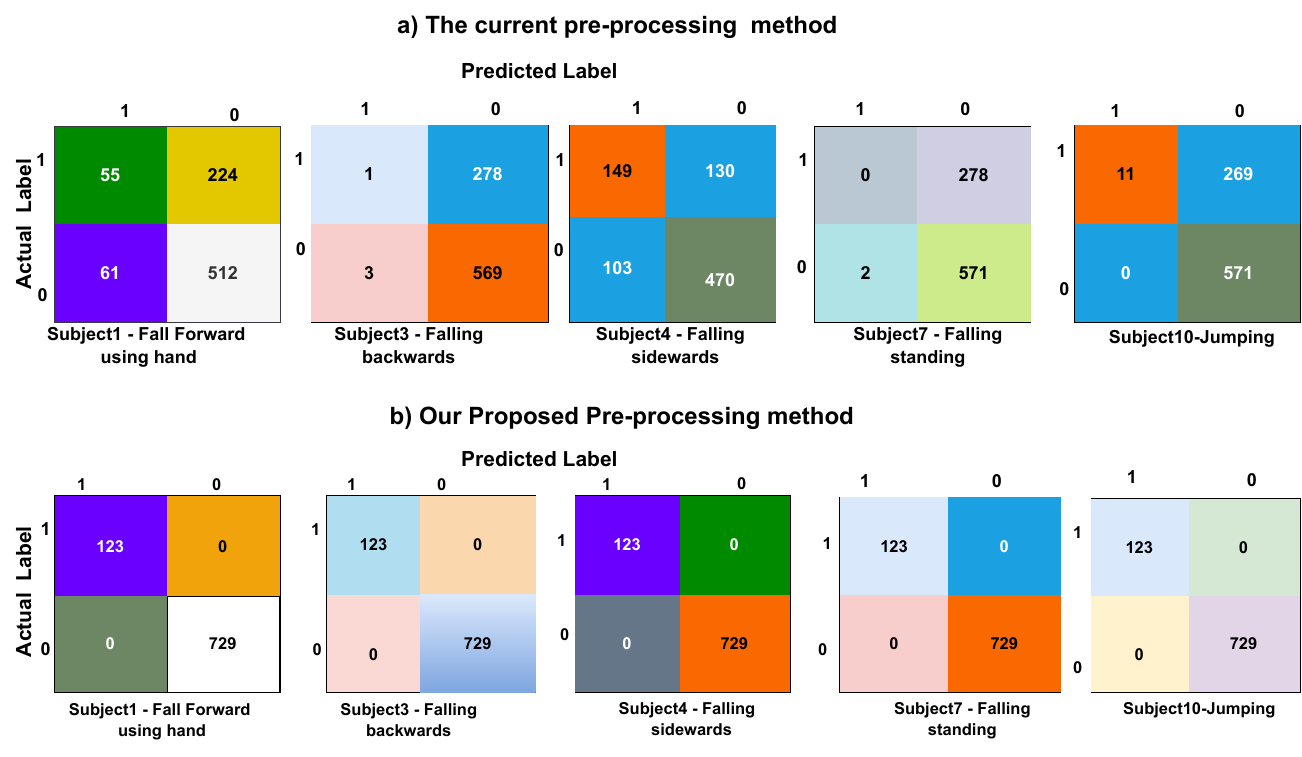}
    \vspace{-5pt} % Adjust the vertical space as needed
    \caption{Confusion Matrix comparison: a) current pre-processing \cite{al2021towards} trained with Random Forest; b) our proposed pre-processing trained with Random Forest Algorithm}
    \label{confusion}
\end{figure}

The Confusion matrix depicted in Fig. \ref{confusion} showcases the state-of-the-art processing technique and our proposed preprocessing technique, both using random forest algorithm. The comparison of the two confusion matrices allows us to assess the effectiveness of our proposed preprocessing method. By analyzing the true positive and true negative rates, we can define that our approach achieves better accuracy in detecting falls and non-fall events across different subjects and fall types.

\subsection{FEATURES SELECTION}

Feature selection is a crucial step in solving
classification problems. It aims to reduce the dimensionality of the feature space by selecting the most relevant features for a specific task, such as identifying impacts within fall events.

While various feature selection methods, such as statistical techniques, have been utilized in previous studies \cite{chandra2011efficient}. However, for this particular work, an embedded feature selection within the machine-learning approach will be adopted to select the significant features for our model's training. This approach offers the advantage of leveraging the power of machine learning algorithms to automatically identify the most important features of the task at hand.

\begin{itemize}

\item \textbf{Selection of the best Feature using Random forest Algorithm}\end{itemize}
The random forest algorithm is well-suited for feature selection due to its ability to calculate feature importance based on the Gini impurity measure.
By utilizing the random forest algorithm for feature selection, the importance of each feature can be assessed and ranked based on the important value of Gini impurity, which is calculated by Equation. 4:

\begin{equation}
\text{Gini impurity} = 1 - \sum_{i} (p_i)^2 \tag{4}
\end{equation}

where \(p_i\) represents the proportion of samples belonging to class \(i\) in the subset of data associated with a particular feature.

This provides valuable insights into which features have the greatest impact on accurately identifying impacts in fall events.
\begin{figure}[t]
    \centering
    \includegraphics[width = 0.50\textwidth]{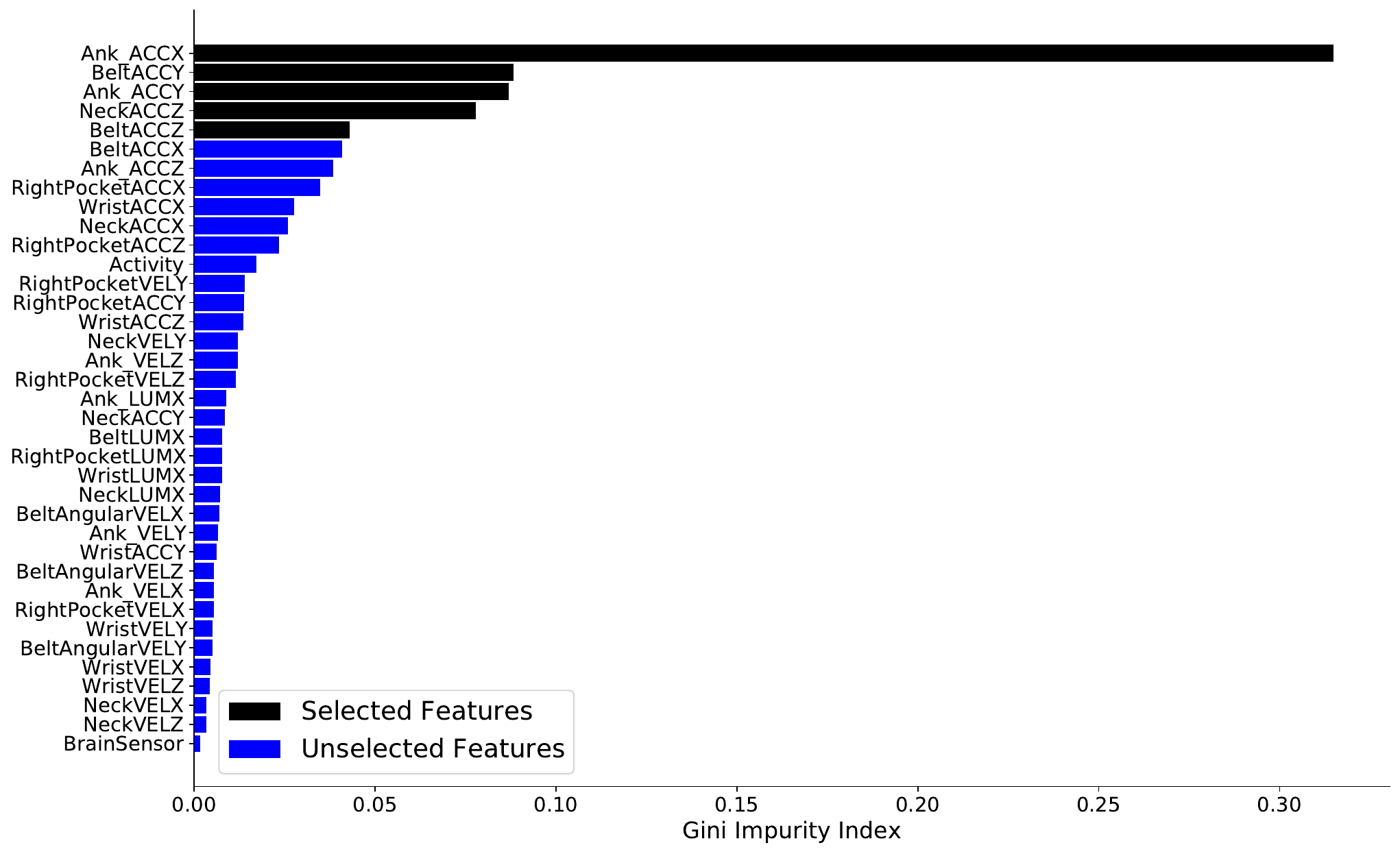}
    \caption{Feature Ranking}
    \label{fig:my_label}
\end{figure}

Thus, based on the GINI impurity method, we first rank the features as in Fig. \ref{fig:my_label}. Next, as shown in TABLE \ref{comparison}, after a thorough evaluation of  the comparison of the time-consuming and the Out of Bag (OOB) error metric  \cite{janitza2018overestimation}  between the top 3, top 5, top 10 features, top 20, and all the features. we select the top five (5) best-ranked features for our impact detection model \ref{fig:my_label}.

\begin{table}[t]
\centering
\caption{Comparison of the training time and the oob error rate. The top 5 in bold and its corresponding numerical values refer to significant features selected.}

\label{comparison}

\scalebox{1.2}{
\begin{tabular}{lcccr}
\hline
Feature selected & OOB Error rate & \ Training Time (s)\\ 
\hline
Top 3 & 0.011 & 4.69 \\
\textbf{Top 5} & \textbf{0.008} & \textbf{6.11} \\
Top 10 & 0.007 & 8.82 \\
Top 20 & 0.006 & 10.92 \\
All features & 0.007  & 15.83 \\
\hline
\end{tabular}}
\end{table}
Choosing the top five important features is a balanced decision. While the top three features have a lower OOB error rate and training time, selecting only a few features can result in the loss of valuable information from the discarded features. This may reduce the model's predictive performance or increase bias if the discarded features are relevant. On the other hand, selecting more features like the top 10, 20, or 37 can increase training times and the risk of overfitting. Hence, By opting for the top five features, we reduce the complexity of the model while still capturing important information. Moreover, it allows for a better understanding of the relationships between the features and the target variable. This choice can enhance the model's generalization, making it more robust and less prone to overfitting on irrelevant or noisy features. Therefore choosing the top 5 features as important features for our  models will be more efficient and effective to detect the impact within fall events.

\subsection{IMPACT 
DETECTION USING MACHINE LEARNING}
In our study, we employed eight different machine learning models for impact point identification within fall events. These models were implemented using Python 3.6 and were run on a PC with the following specifications: Intel Core 32GB RAM, and a graphic card GeForce 3080Ti with 16GB RAM. Each machine learning model was trained and evaluated using appropriate performance metrics such as accuracy, precision, recall, and F1-score, taking into consideration their respective hyperparameters as shown in 
Table \ref{tab:hyperparameters}.
These metrics provide a comprehensive evaluation of the models' performance in  impact detection within fall events.

\begin{table}[t]
\centering
\caption{ALGORITHMS TRAINING’ HYPERPARAMETERS}
\label{tab:hyperparameters}
\begin{tabularx}{\columnwidth}{lX}
\hline
\textbf{Algorithm} & \textbf{Hyperparameters} \\
\hline
SVM & \{'C': 1.0, 'kernel': 'rbf', 'gamma': 'scale'\} \\
Logistic Regression & \{'C': 1.0, 'solver': 'lbfgs'\} \\
Decision Tree & \{'criterion': 'gini', 'max\_depth': None\} \\
K-Nearest Neighbors & \{'n\_neighbors': 5, 'weights': 'uniform'\} \\
Gaussian Naive Bayes & - \\
Random Forest & \{'n\_estimators': 100, 'criterion': 'gini'\} \\
Stochastic Gradient Descent & \{'alpha': 0.0001, 'max\_iter': 1000\} \\
Gradient Boosting & \{'n\_estimators': 100, 'learning\_rate': 0.1\} \\
\hline
\end{tabularx}
\end{table}

\begin{itemize}\item \textbf{SVM}\end{itemize}
Support Vector Machine (SVM)\cite{noble2006support} is  commonly used for  fall detection system\cite{liu2012fall}. By using the SVM we aim to capitalize on its ability to capture complex patterns and distinguish impact within fall events.

\begin{itemize}\item \textbf{Decision Tree}\end{itemize}
In the context of impact detection in fall events, DTs\cite{song2015decision} can be utilized to accurately identify the moment of impact. By constructing a decision tree based on relevant features and conditions associated with impact, the model can effectively classify instances as either impact or non-impact.

\begin{itemize}\item \textbf{Random Forest (RF)}\end{itemize}

In the context of fall and impact detection, RF\cite{paul2018improved} can effectively leverage the strengths of DTs to identify impact moments accurately. By considering multiple DTs trained on different subsets of data, RF can handle complex relationships and capture important features for impact detection.
\begin{itemize}\item \textbf{Gradient Boosting (GB)}\end{itemize}
In the context of fall and impact detection, GB\cite{ke2017lightgbm} can be a suitable algorithm due to its ability to handle sequential learning and refine the impact detection process over iterations.
\begin{itemize}\item \textbf{SGD Classifier}\end{itemize}

The SGD\cite{gunale2018indoor} learns to distinguish between the two classes based on the input features provided. By utilizing gradient descent optimization, the classifier adjusts its parameters to minimize the classification error and improve its ability to correctly identify impacts.
\begin{itemize}\item \textbf{Naive Bayes}\end{itemize}
In the context of impact detection in fall detection systems, Naive Bayes\cite{altay2019use} can be a viable choice due to its simplicity, efficiency, and ability to handle high-dimensional data.

\begin{itemize}\item \textbf{Linear Regression}\end{itemize}
By utilizing LR \cite{putra2017event} for impact detection in fall events, it is possible to leverage its probabilistic nature and make informed predictions to enhance the accuracy and reliability of impact and fall detection systems.

\section{EXPERIMENTAL RESULTS}\label{sec:results}

In this section, we present the results of our impact detection models and provide a detailed discussion of their performance. The evaluation of the models was carried out using various metrics to assess their effectiveness in accurately detecting impacts in fall events.

The evaluation protocol involved dividing the dataset into three subsets: a training set, a validation set, and a test set. The training set, comprising 80\% of the dataset, was used to train the models, while the validation set (10\% of the dataset) and the test set (10\% of the dataset) were used for performance assessment respectively. TABLE \ref{tab:performance_metrics} below presents the key findings of our proposed methods on test dataset. 

\begin{table}[t]
\centering
\caption{Machine Learning Metrics on Impact Detection tasks. Bold refers to the highest values}
\label{tab:performance_metrics}
\begin{tabularx}{\columnwidth}{l*{5}{X}}
\hline

\textbf{Algorithm} & \textbf{Accuracy (\%)} & \textbf{Recall (\%)} & \textbf{Precision (\%)} & \textbf{F1-Score (\%)} & Training Time (s)\\
\hline
SVM & \textbf{99.50} & \textbf{99.50} & \textbf{99.50} & \textbf{99.50} & 0.059 \\
RF & \textbf{99.28} & \textbf{98.47} & \textbf{99.18} & \textbf{98.47} & 0.607\\
SGD & 94.47 & 97.71 & 95.85 & 96.77 & 0.013 \\
NB & 98.85 & 97.57 & 97.94 & 97.76 & 0.001\\
DT & 98.85 & 97.57 & 97.94 & 97.76 & 0.034\\
KNN & 98.35 & 98.45 & 99.03 & 98.74 & 0.005\\
LR & 95.75 & 87.71 & 91.20 & 94.64 & 0.010\\
GBOOST & \textbf{99.35} & \textbf{98.65} & \textbf{98.84} & \textbf{98.74} & 0.844 \\
\hline
\end{tabularx}
\end{table}
In summary, our results highlight the effectiveness of various machine-learning models in detecting the impacts in fall events. The high recall and accuracy values achieved by the SVM, GB, and RF models suggest that they are capable of accurately detecting impact points and minimizing false negatives. These models can provide timely assistance and support to individuals in need, reducing the risk of severe injuries and complications associated with falls.
\\

\begin{figure}[t]
    \centering
    \includegraphics[width = 0.46\textwidth]{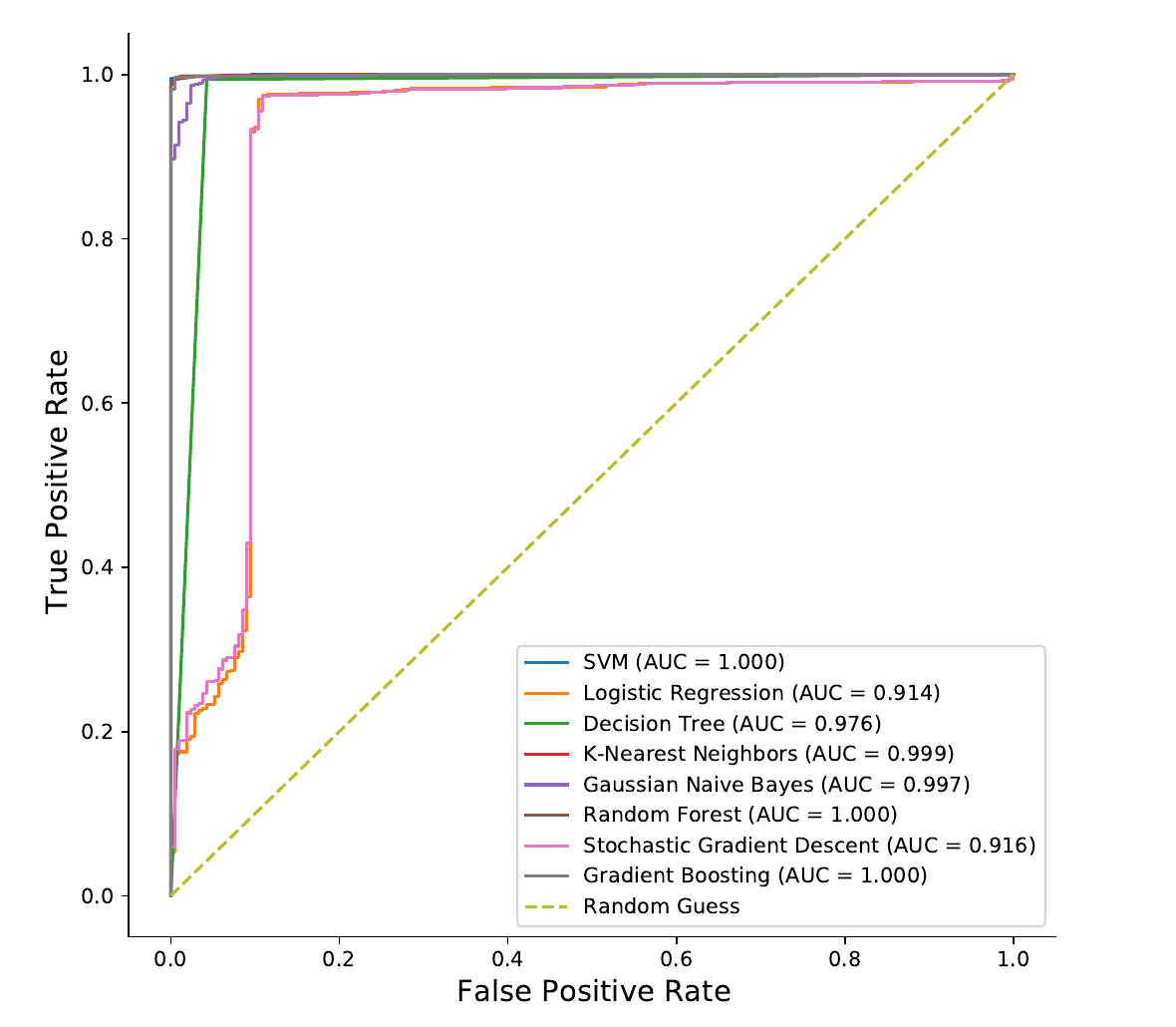}
    \caption{ROC curves for the  Machine learning Models used}
    \label{roc}
\end{figure}

The ROC curve in Fig.\ref{roc} provides a numerical value of AUC that quantifies the performance of our algorithms used in this work.
% AUC is a measure of the overall performance of a binary classification model, and a value of 1 indicates a perfect classifier. Therefore, the SVM, Random Forest, and Gradient Boosting models, with an AUC value of 1, exhibit good performance in accurately detecting impacts within fall events. This shows that these models can reliably differentiate between true positives and false positives, enhancing their effectiveness in fall detection systems.
It measures the overall performance of a binary classification model, and a value of 1 indicates a perfect classifier. This indicates that the models with such score can reliably differentiate between true positives and false positives.

However, despite the SVM model exhibiting a good AUC score, it has a shorter inference time compared to the other models. This means that the SVM model can make predictions more quickly when deployed in a real-time or time-sensitive scenario.  Thus, with the combination of high AUC values and the related inference time, the SVM model is particularly effective in accurately identifying the impact within fall events.
 
\section{CONCLUSION}\label{sec:conclusion}

In conclusion, this study highlights the significance of appropriate pre-processing techniques in enhancing the accuracy and reliability of impact detection in fall events. The evaluation of various machine learning models revealed SVM as the most promising model for impact detection due to its high accuracy, robustness, and shorter inference time. These qualities make SVM suitable for fall impact identification in real-time or time-sensitive scenarios.
The findings of this study contribute to the advancement of fall detection technologies by highlighting the significance of pre-processing techniques and identifying effective machine learning models for impact detection.
Moving forward, future work should focus on exploring the integration of multimodality by combining data from different sources. Additionally, the proposed pre-processing technique should be applied to new datasets to assess its performance and improve the accuracy, reliability, and adaptability of impact and fall detection systems.

\bibliographystyle{IEEEtran}
\bibliography{reference.bib}

\end{sloppypar}
\end{document}